\documentclass[prl,twocolumn,amsmath,amssymb,superscriptaddress,floatfix]{revtex4-2}
 
\usepackage[english]{babel}
\usepackage{amsfonts}
\usepackage{graphicx}
\usepackage{times}
\usepackage{color}
\begin{document}
 
\title{Hallmark Signatures of Electronic Pairing in
       Two-Photon Two-Electron Coincidence
       Angle-Resolved Photoemission Spectroscopy}
 
\author{Janez \surname{Bon\v ca}}
\affiliation{Faculty of Mathematics and Physics, University of Ljubljana,
  1000 Ljubljana, Slovenia}
\affiliation{J.\ Stefan Institute, 1000 Ljubljana, Slovenia}
 
\author{Alberto Nocera}
\affiliation{Department of Physics and Astronomy, University of British
  Columbia, Vancouver, BC, Canada V6T~1Z1}
\affiliation{Quantum Matter Institute, University of British Columbia,
  Vancouver, BC, Canada V6T~1Z4}
 
\author{Andrea \surname{Damascelli}}
\affiliation{Department of Physics and Astronomy, University of British
  Columbia, Vancouver, BC, Canada V6T~1Z1}
\affiliation{Quantum Matter Institute, University of British Columbia,
  Vancouver, BC, Canada V6T~1Z4}
 
\author{Mona \surname{Berciu}}
\affiliation{Department of Physics and Astronomy, University of British
  Columbia, Vancouver, BC, Canada V6T~1Z1}
\affiliation{Quantum Matter Institute, University of British Columbia,
  Vancouver, BC, Canada V6T~1Z4}
 
\date{\today}
 
\begin{abstract}
Understanding strongly correlated quantum materials remains a central challenge in condensed matter physics and materials science.  While angle-resolved photoemission spectroscopy (ARPES) has become an indispensable probe of single-quasiparticle excitations, it accesses electronic correlations only indirectly.  Here we show that unlike one-photon in, two-electrons out coincidence ARPES ($\gamma\!\rightarrow\!2e$ 2eARPES), the two-photon in, two-electron out $2\gamma\!\rightarrow\!2e$ 2eARPES provides a direct and unambiguous probe of electronic pairing.  We establish this on general theoretical grounds and substantiate it through large-scale numerical simulations of strongly correlated models with both paired and unpaired ground states.  The key result is a model-independent separation in the $(\omega_1,\omega_2)$ plane of the two photoelectrons' energies, between signal from electrons emitted from the \emph{same} pair and signal from electrons emitted from \emph{different} pairs; this follows from energy conservation alone and is independent of any material-specific assumptions.  Our findings demonstrate that $2\gamma\!\rightarrow\!2e$ 2eARPES can identify pairing and extract the pair binding energy as well as the energy of the 'glue' boson without any sophisticated data analysis or complementary measurements.
\end{abstract}
 
\maketitle
 
\textit{Introduction:} Quantum materials exhibit a wealth of emergent properties that make them central platforms for future technologies.  Advancing our understanding of these phenomena—and learning to control and engineer them through materials design—requires detailed insight into the physics of strongly correlated electron systems.
 
Significant theoretical progress has been made in this direction, particularly following the discovery of high-temperature superconductivity in the cuprates.  Experimentally, a growing array of spectroscopic techniques now provides unprecedented access to the properties of quantum materials.  Nevertheless, a complete microscopic understanding of why these materials behave as they do remains elusive.
 
Part of the difficulty is that current experimental probes measure correlations only indirectly.  For example, ARPES measures single-quasiparticle properties with great precision, yet the effects of correlations manifest only through renormalizations of the quasiparticle dispersion and spectral weight, which require substantial theoretical modeling to interpret~\cite{Damascelli2003}.
 
Motivated by this limitation, new experimental approaches aimed at directly probing two-particle correlations are being proposed and evaluated~\cite{Su2025n}.  Among these, two-electron coincidence ARPES (2eARPES) has attracted considerable attention and substantial investment from funding agencies for the construction of next-generation instruments~\cite{Boschini2024}.
 
Two distinct 2eARPES setups have been proposed: $\gamma\!\rightarrow\!2e$ (one photon in, two electrons out) and $2\gamma\!\rightarrow\!2e$ (two photons in, two electrons out). Theoretical work has been devoted to both~\cite{haak1978,gunnarsson1981,berakdar1998,napitu2010,Su2020,Mahmood22,Devereaux_2023,Kemper2025,Su2025n,old}, while experimental results are so far available only for the former~\cite{Kirschner_1998,Kischner2014,Kischner2017,Kirschner2007, Kischner2020}.  In both configurations, photoelectrons are detected in coincidence with the goal of extracting information about their correlations from the joint intensity.  For systems with paired electrons, the ability to distinguish the signal from electrons emitted from the same pair from the signal due to electrons emitted from different pairs should confirm pairing and give access to pair properties.      A central question is therefore to what extent these two contributions can be cleanly separated and whether the properties of the pairs can be directly extracted from the resulting signals.

\begin{figure}[b]
\includegraphics[width=\columnwidth]{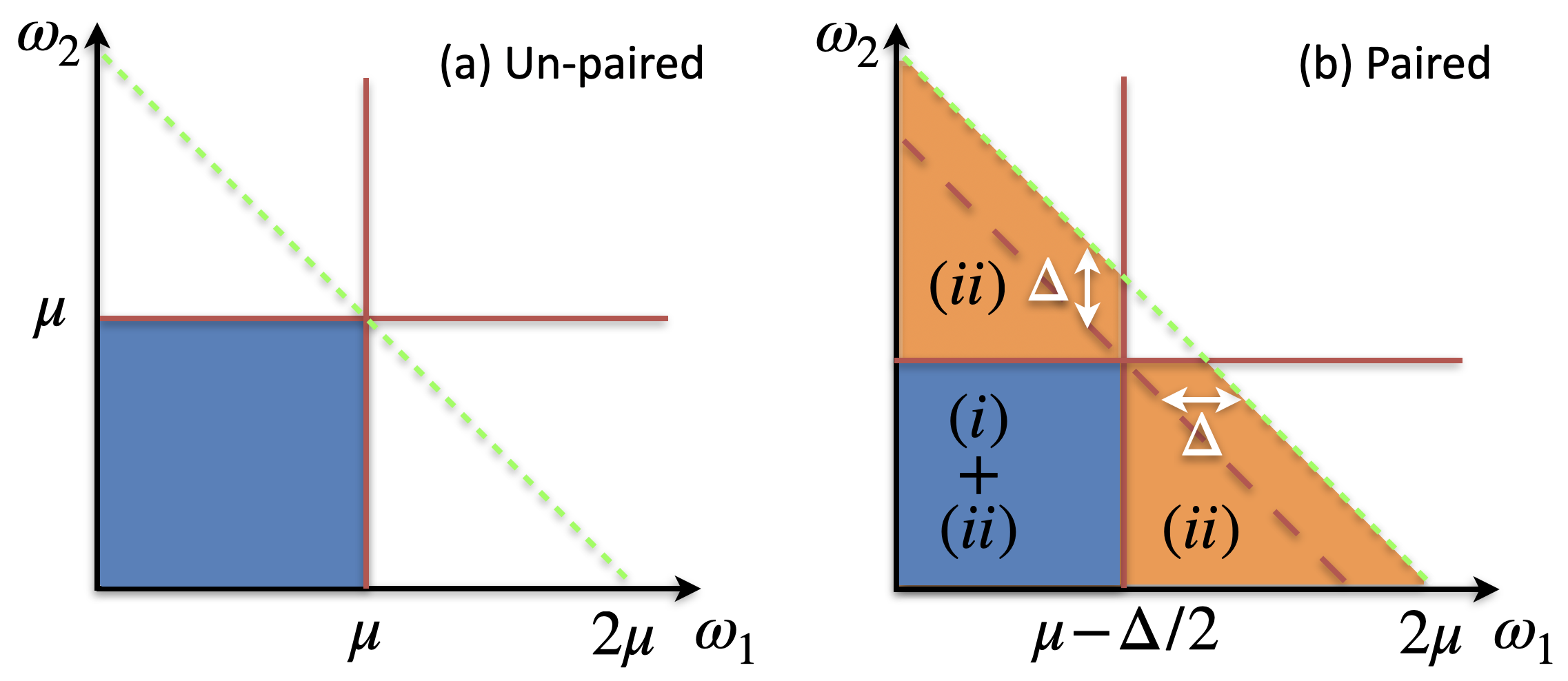}
\caption{\label{FIG1}
  Shaded regions indicate where $2\gamma\!\rightarrow\!2e$ 2eARPES  spectral weight can appear in the $(\omega_1,\omega_2)$ plane of the photoelectrons' energies, for (a)~a quasiparticle liquid and (b)~a liquid of pairs with binding energy $\Delta$; $\mu$ is the chemical potential.  In (b), the orange  regions labeled `(ii)' are accessible only when both electrons  originate from the same pair, while the blue region admits contributions from electrons ejected from different pairs, labeled `(i)'.  The significance of the solid and dashed lines is explained in the text.}
\end{figure}

In this Letter, we show that when resolved in the individual energies $(\omega_1,\omega_2)$ of the two coincident photoelectrons, the $2\gamma\!\rightarrow\!2e$ signal from electrons ejected from \emph{different} pairs [hereafter type-(i) processes] is spectrally separated from that due to electrons ejected from the \emph{same} pair [type-(ii) processes].  This separation is a direct consequence of energy conservation and gives rise to qualitatively distinct intensity patterns in paired versus unpaired systems,  summarized in Fig.~\ref{FIG1}.  The pair binding energy is directly readable from the $2\gamma\!\rightarrow\!2e$ intensity map, while finer details of the pair wavefunction are accessible with further analysis. Our results constitute a major advance because they demonstrate that this distinction is model-independent, thereby establishing $2\gamma\!\rightarrow\!2e$ 2e-ARPES as a powerful spectroscopic tool capable of unambiguously identifying pairing and directly extracting the binding energy without the need for advanced data analysis or detailed theoretical modeling. These findings are consistent with, while substantially extending and unifying within a common framework, the results reported in Refs.~\cite{Devereaux_2023,old}.

In contrast, in the $\gamma\!\rightarrow\!2e$ configuration this clear difference between the energy-resolved spectra for paired and unpaired systems is masked by overlap with the signal from regular ARPES processes, as discussed in the End Matter.  While this setup may offer advantages for probing other aspects of electronic correlations, our focus here is on paired systems; accordingly, we discuss only the $2\gamma\!\rightarrow\!2e$ setup hereafter, referring to it simply as 2eARPES.

\textit{Universal bounds for 2eARPES: } The momentum- and energy-resolved coincident photoelectron intensity in the $2\gamma\!\rightarrow\!2e$ configuration is  proportional to~\cite{Su2020,Devereaux_2023}:
\begin{widetext}
\begin{align}
D_{{\bf k}_1,{\bf k}_2}(\omega_1,\omega_2)
  = \sum_{n}
     \bigl|M^{(n)}_{{\bf k}_1,{\bf k}_2}(\omega_1,\omega_2)\bigr|^2
     \delta\!\left(
       E_{-{\bf k}_1-{\bf k}_2}^{(n,N_e-2)}
       - E_{\bf 0}^{(0,N_e)}
       + \omega_1 + \omega_2
     \right). \label{Domom}
\end{align}
\end{widetext}
Here $\omega_i = {\bf K}_i^2/2m + W - \omega_{\rm ph}$, $i=1,2$, are the energies removed from the system upon absorption of a photon of energy $\omega_{\rm ph}$ (we set $\hbar=1$) and emission of a photoelectron with momentum ${\bf K}_i$; $W$ is the work function. The components of the crystal momenta ${\bf k}_i$ parallel to the sample surface are conserved: ${\bf K}_{i,\parallel}={\bf k}_{i,\parallel}$. The eigenstates and eigenenergies of $\mathcal{H}$ are ${\cal H}|\psi_{\bf k}^{(n,N_e)}\rangle  = E_{\bf k}^{(n,N_e)}|\psi_{\bf k}^{(n,N_e)}\rangle$, where $N_e$ is the electron number, ${\bf k}$ the total crystal momentum, and $n=0$ ($n\geq 1$) labels the ground state (excited states) at fixed $N_e$.  Thus the  $N_e$-electron ground state  $|\psi_{\bf 0}^{(0,N_e)}\rangle$ has energy $E_{\bf 0}^{(0,N_e)}$.
 
The $\delta$-function in Eq.~(\ref{Domom}) encodes global energy conservation for the $2\gamma\!\rightarrow\!2e$ process.  Its counterpart in conventional ARPES~\cite{Damascelli2003}, $\delta(E_{-{\bf k}}^{(n,N_e-1)}-E_{\bf 0}^{(0,N_e)}+\omega)$, requires $\omega = E_{\bf 0}^{(0,N_e)}-E_{-{\bf k}}^{(n,N_e-1)} \leq E_{\bf 0}^{(0,N_e)}-E_{\bf 0}^{(0,N_e-1)} = \mu$, so ARPES spectral weight can appear only below the chemical potential. An analogous argument applied to Eq.~(\ref{Domom}) yields the global constraint (dashed green line in Fig.~\ref{FIG1}):
\begin{equation}
    \omega_1 + \omega_2 \leq 2\mu.
    \label{global}
\end{equation}
 
The matrix element in Eq.~(\ref{Domom}) is given by~\cite{Su2020,Devereaux_2023}:
\begin{widetext}
\begin{equation}
M^{(n)}_{{\bf k}_1{\bf k}_2}(\omega_1,\omega_2) =
\sum_{m\sigma_1\sigma_2} \left[
\frac{
  \langle\psi_{-{\bf k}_1-{\bf k}_2}^{(n,N_e-2)}|
    c_{{\bf k}_2\sigma_2}|
    \psi_{-{\bf k}_1}^{(m,N_e-1)}\rangle\,
  \langle\psi_{-{\bf k}_1}^{(m,N_e-1)}|
    c_{{\bf k}_1\sigma_1}|
    \psi_{\bf 0}^{(0,N_e)}\rangle
}{
  E_{{\bf k}_1}^{(m,N_e-1)} - E_{\bf 0}^{(0,N_e)} + \omega_1 - i\eta
}
- (1\leftrightarrow 2)
\right], \label{Mkk}
\end{equation}
\end{widetext}
where the second term is obtained from the first by exchanging all labels $1\leftrightarrow 2$.
 
This expression imposes additional restrictions on the region of the $(\omega_1,\omega_2)$ plane where 2eARPES intensity can appear, beyond the global constraint~(\ref{global}).  Our goal is to contrast these regions for paired and unpaired systems.  To maintain maximum generality, we focus on the momentum-integrated 2eARPES intensity:
\begin{align}
D(\omega_1,\omega_2)
  = \frac{1}{N^2}\sum_{{\bf k}_1,{\bf k}_2}
    D_{{\bf k}_1,{\bf k}_2}(\omega_1,\omega_2). \label{Dsum}
\end{align}
This choice is advantageous for several reasons: (a)~As argued below and illustrated in Fig.~\ref{FIG1}, differences in the allowed $(\omega_1,\omega_2)$ regions with non-vanishing $D(\omega_1,\omega_2)$ suffice to unambiguously distinguish paired from unpaired ground states, for any Hamiltonian $\mathcal{H}$ in any spatial dimension.  (b)~These same constraints remain valid at any fixed momenta $({\bf k}_1,{\bf k}_2)$. (c)~Momentum integration enables more efficient experimental data collection.
 
The denominators in Eq.~(\ref{Mkk}) impose two further constraints:
\begin{align}
    \omega_i = E_{\bf 0}^{(0,N_e)} - E_{-{\bf k}_i}^{(m,N_e-1)},
    \quad i=1,2.
\end{align}
 
Consider first a normal metal, whose low-energy excitations are weakly interacting quasiparticles.  By the same reasoning as for ARPES, the denominators require $\omega_i\leq\mu$ for $i=1,2$.  Since the two electrons are photoemitted independently, both constraints must be satisfied simultaneously, confining 2eARPES spectral weight to the region $\omega_1\leq\mu$, $\omega_2\leq\mu$ [Fig.~\ref{FIG1}(a)].  The global constraint~(\ref{global}) is automatically satisfied in this case.
 
We now turn to paired ground states, which we model as collections of weakly interacting  singlets.  This encompasses the ground state of most superconductors.  In one-dimensional systems, paired ground states without long-range order arise due to quantum fluctuations; a prominent example is the bipolaron liquid of the 1D Hubbard-Holstein model at low carrier densities~\cite{Klemen2025}.  Although our analysis is strictly at zero temperature, we expect the conclusions to extend naturally to liquids of preformed pairs hypothesized to exist above the superconducting critical temperature $T_C$ and below the pair-dissociation temperature $T^*$.  This regime is relevant to superconductors in which $T_C$ marks the loss of phase coherence due to phase fluctuations, rather than thermal pair breaking as in conventional BCS superconductors~\cite{BCS1,BCS2}.
 
As detailed below, the shaded regions in Fig.~\ref{FIG1}(b) show the extended region where 2eARPES spectral weight appears when the pair binding energy is $\Delta$.  The global constraint~(\ref{global}) (dashed green line) continues to hold. Unlike in the unpaired case, however, weight can extend all the way to this boundary—except within the central right-angled  triangle of leg length $\Delta$, where weight necessarily vanishes. 

The contrast between panels (a) and (b) is striking and provides an unambiguous spectroscopic signature distinguishing the two ground states.  This pronounced difference arises because the paired ground state admits the two qualitatively distinct contributions to $D(\omega_1,\omega_2)$: photoelectrons originate from \emph{different} pairs or from the \emph{same} pair.
 
To establish energy bounds for each contribution, we approximate the ground-state energy for weakly interacting pairs as $E_{\bf 0}^{(0,N_e)} \approx 2E_{\rm pair} + E_{\bf 0}^{(0,N_e-4)}$, where $E_{\rm pair} = 2E_{\rm qp} - \Delta$ is the minimum pair energy, $\Delta$ is the binding energy, and $E_{\rm qp}$ is the minimum energy of an unbound quasiparticle (qp).  Upon ejection of one electron from a pair, the remaining unpaired electron contributes at least $E_{\rm qp}$ to the total energy.  For type-(i) processes, breaking two separate pairs leaves behind at least two quasiparticles, giving $E_{-{\bf k}_1-{\bf k}_2}^{(n,N_e-2)}  \geq 2E_{\rm qp} + E_{\bf 0}^{(0,N_e-4)}$. Here we invoked the customary \emph{sudden} approximation, in which the photoelectrons depart before the remaining electrons can relax by re-pairing.  Substituting into the global constraint yields $\omega_1+\omega_2 \leq 2E_{\rm pair}-2E_{\rm qp} = E_{\rm pair}-\Delta$, shown by the dashed red line.  For type-(ii) processes, in which both electrons are emitted from the same pair, the entire pair is removed, giving $E_{-{\bf k}_1-{\bf k}_2}^{(n,N_e-2)}  \geq E_{\rm pair} + E_{\bf 0}^{(0,N_e-4)}$, and the global constraint gives $\omega_1+\omega_2 \leq 2E_{\rm pair}-E_{\rm pair}= 2\mu$, shown by the dashed green line.  The gap $\Delta$ between these two bounds reflects the additional energy cost of breaking a second pair, as already noted in Ref. \cite{old}.
 
The denominators in Eq.~(\ref{Mkk}) impose further constraints. Emission of the first electron leaves a quasiparticle in an intermediate $(N_e\!-\!1)$-electron state with $E_{-{\bf k}_i}^{(m,N_e-1)} \geq E_{\rm qp} + E_{\rm pair} + E_{\bf 0}^{(0,N_e-4)}$, which gives $\omega_i \leq E_{\rm pair}-E_{\rm qp} = \mu-\Delta/2$ (solid red lines).  This coincides precisely with the spectral range of single-particle ARPES weight for a paired ground state~\cite{Klemen2025}.  For type-(i) processes, both $i=1,2$ constraints must hold simultaneously since the two ejections are independent, confining spectral weight to the blue region in Fig.~\ref{FIG1}(b).  For type-(ii) processes, only the \emph{first} ejected electron must satisfy $\omega_i\leq\mu-\Delta/2$; the second can carry all remaining energy from the broken pair.  Consequently, type-(ii) weight requires only $\omega_1\leq\mu-\Delta/2$ \emph{or} $\omega_2\leq\mu-\Delta/2$, in addition to Eq.~(\ref{global}).  This explains both the lack of spectral gap within the central triangle and the distinctive `wing' features (orange regions in Fig.~\ref{FIG1}(b)) accessible to type-(ii) but not to type-(i) processes.

\begin{figure}[t]
\includegraphics[width=0.5\columnwidth]{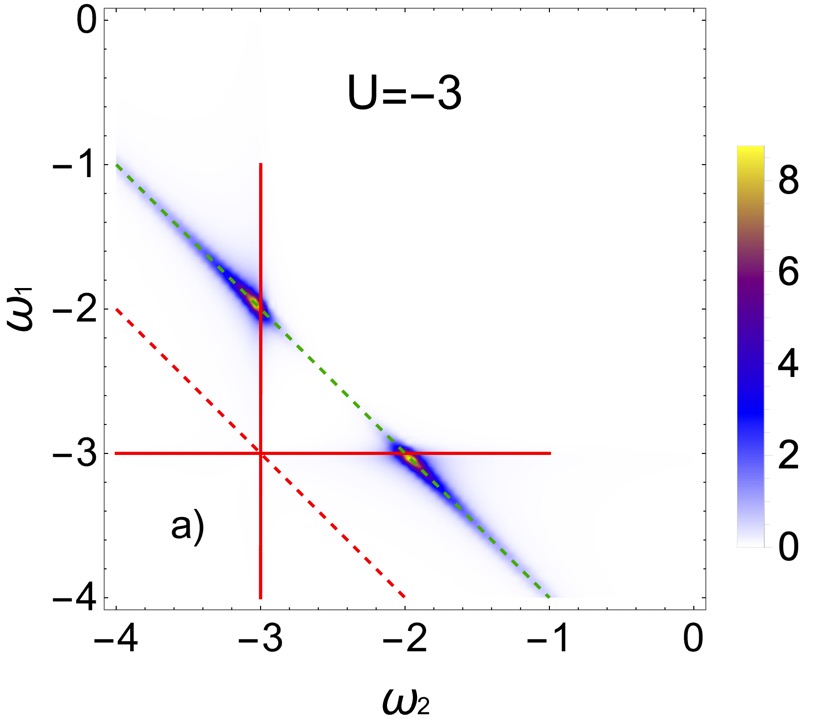}%
\includegraphics[width=0.5\columnwidth]{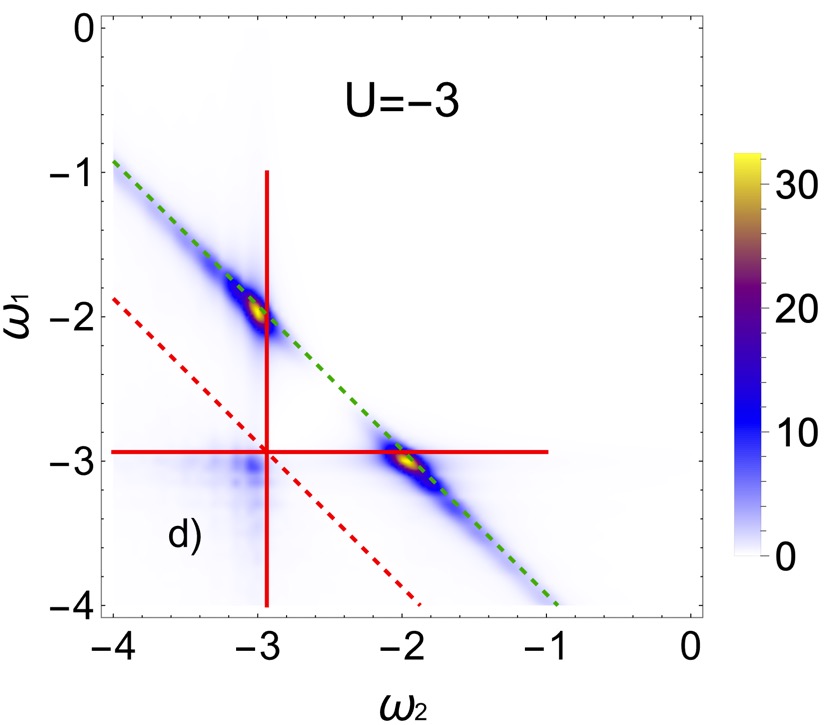}
\includegraphics[width=0.5\columnwidth]{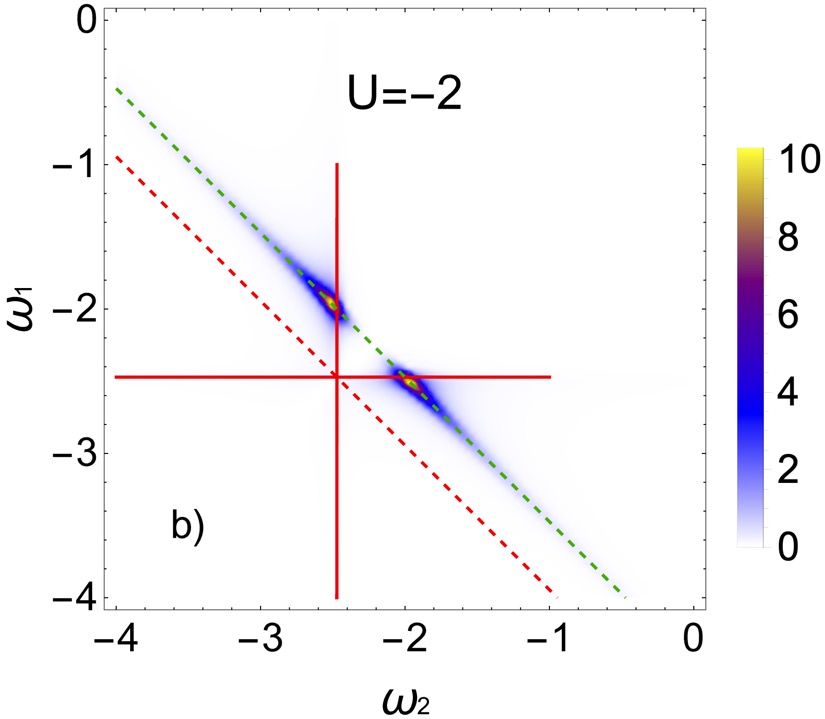}%
\includegraphics[width=0.5\columnwidth]{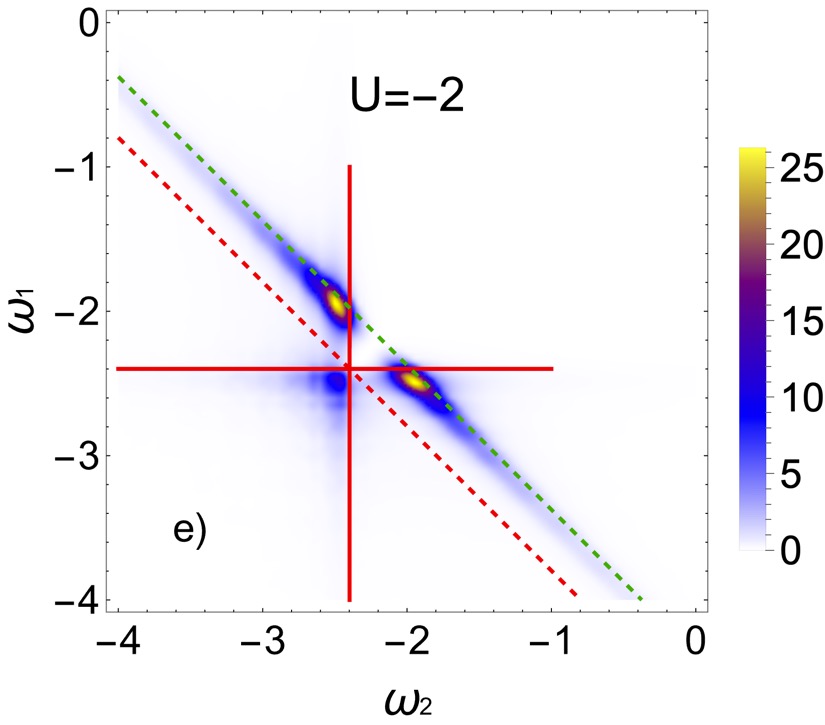}
\includegraphics[width=0.5\columnwidth]{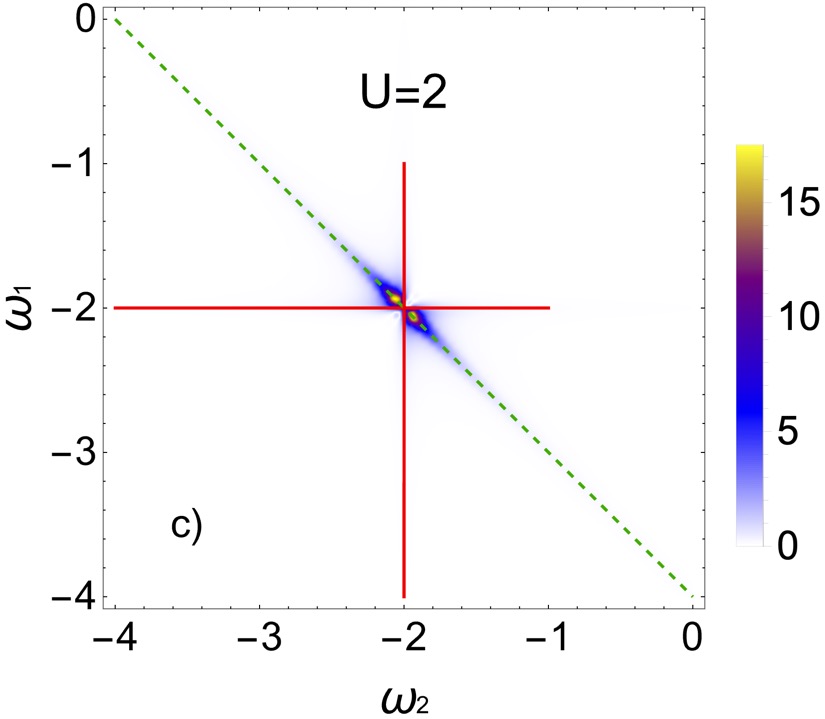}%
\includegraphics[width=0.5\columnwidth]{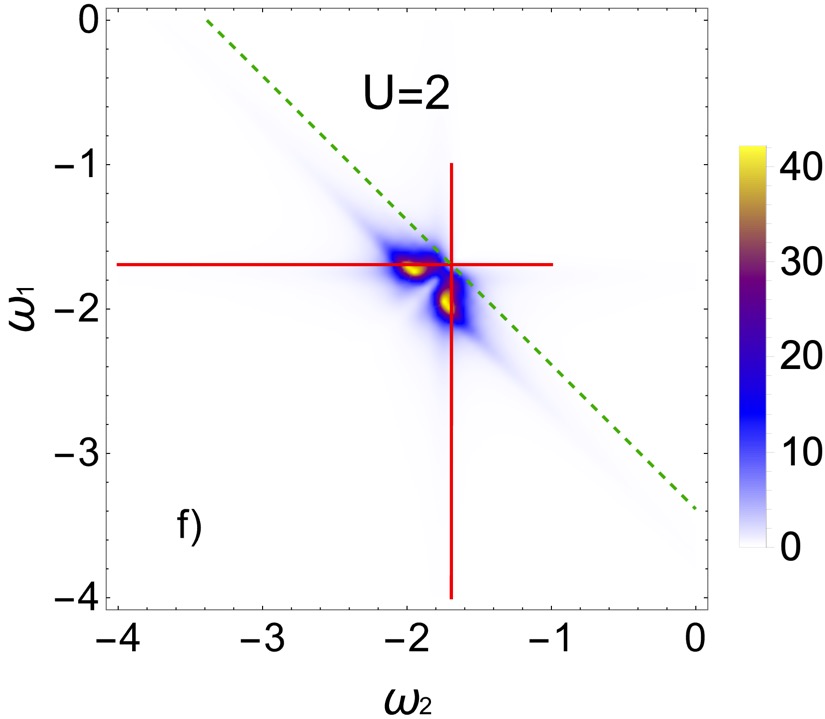}
\caption{\label{FIG2}
  Momentum-integrated 2eARPES weight $D(\omega_1,\omega_2)$ for the pure Hubbard model ($\lambda=0$).  Left column: VED results for  $N_e=2$, $N=90$.  Right column: DMRG results for $N_e=12$, $N=48$ ($n=0.25$).  In all panels $t_{\rm el}=1$, $U=-3,-2,+2$ (top to bottom), and $\eta=0.05$.  Lines as in Fig.~\ref{FIG1}.}
\end{figure}
 
\begin{figure*}
\includegraphics[width=0.49\columnwidth]{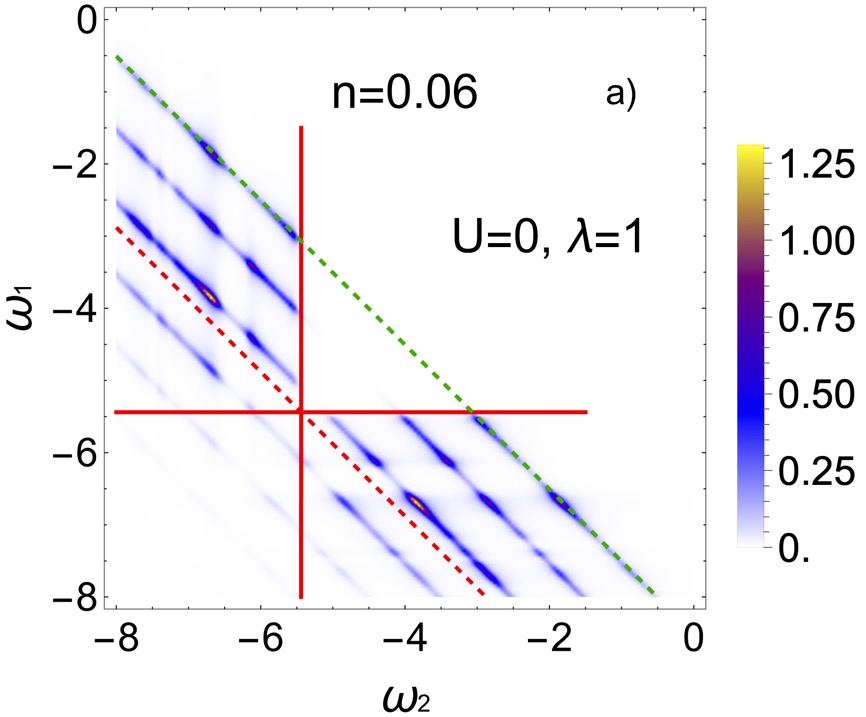}
\includegraphics[width=0.49\columnwidth]{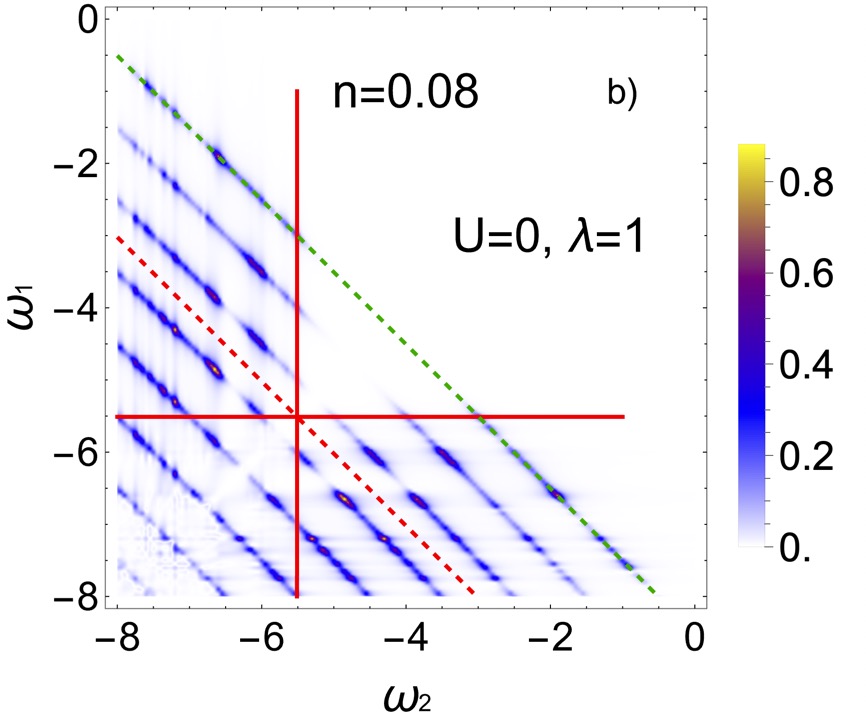}
\includegraphics[width=0.49\columnwidth]{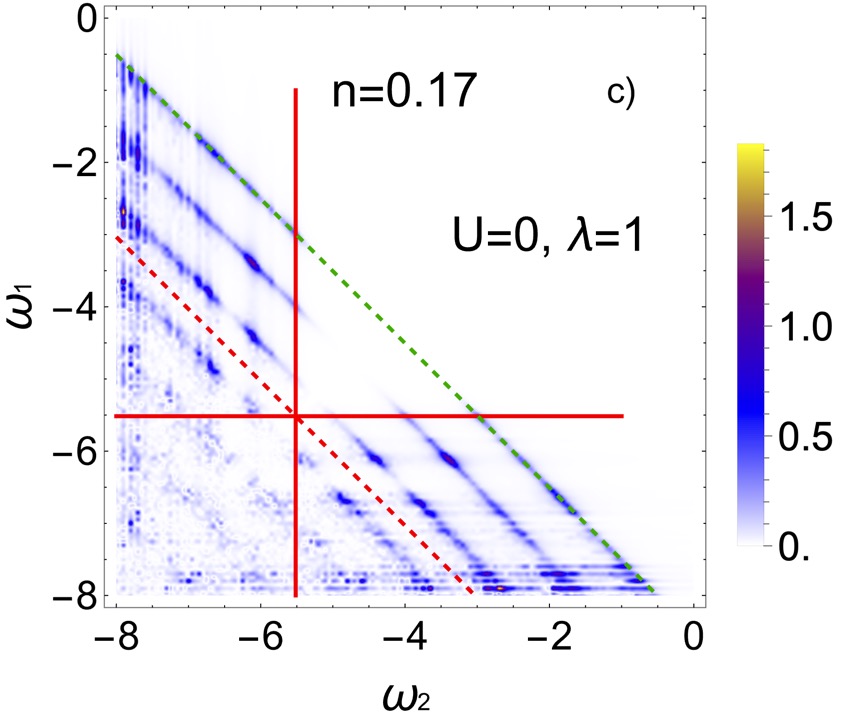}
\includegraphics[width=0.49\columnwidth]{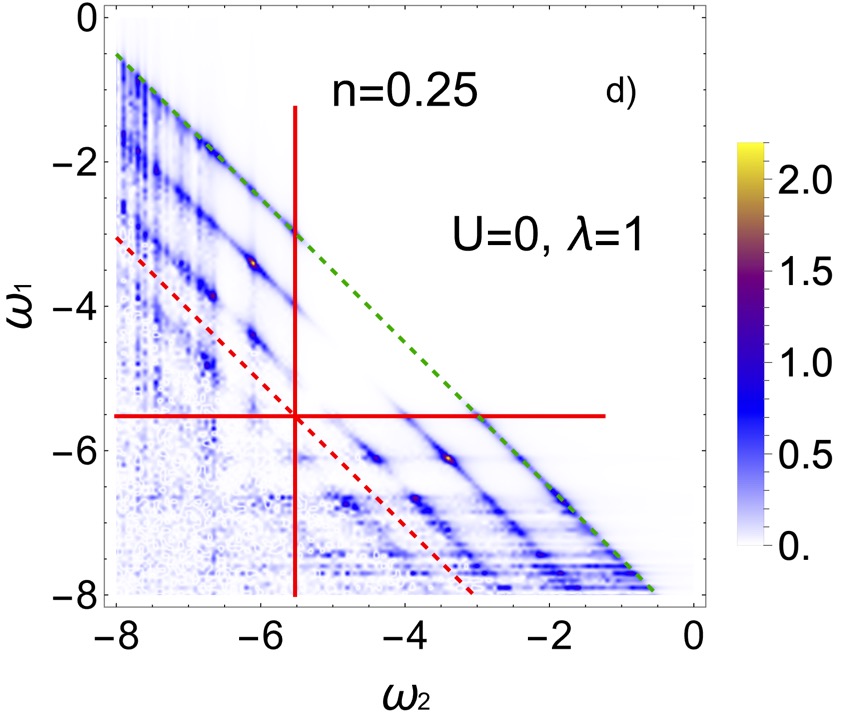}
\caption{\label{FIG3}
  Momentum-integrated 2eARPES weight $D(\omega_1,\omega_2)$ for the  pure Holstein model ($U=0$, $\lambda=1.0$, $\Omega=1.0$, $t_{\rm ph}=0$).  (a)~VED results for $N_e=2$, $N=34$  ($n\to 0$).  (b)--(d)~DMRG results for $N=48$ and $N_e=4,8,12$,  respectively.  Lines as in Fig.~\ref{FIG1}, with $\mu$ and $\Delta$  extracted from DMRG.}
\end{figure*} 
 
These general arguments delineate the \emph{allowed} regions for 2eARPES spectral weight but do not specify its distribution within those regions, which depends on the specific model. Next, we present numerical results for the 1D Hubbard-Holstein model that both verify these universal bounds and reveal the detailed structure of its 2eARPES spectrum.
 
\textit{ Models and methods:} We study an $N$-site chain with lattice constant $a=1$ and $N_e$
electrons, described by the Hamiltonian:
\begin{eqnarray}
&&{\cal H}
  = -t_{\rm el}\sum_{j,\sigma}
      \bigl(c^\dagger_{j,\sigma}c_{j+1,\sigma}+{\rm H.c.}\bigr)
      + U\sum_j n_{j\uparrow}n_{j\downarrow} \nonumber\\
  &&+ t_{\rm ph}\!\!\sum_{j}
      \bigl(b^\dagger_j b_{j+1}\!+\!{\rm H.c.}\bigr)
      \!+ \!\Omega\!\!\sum_j\! b_j^\dagger b_j
      \!+ \!g\!\!\sum_j\hat{n}_j(b_j^\dagger\!+\!b_j)\phantom{aaa}
      \label{ham}
\end{eqnarray}
Here $c^\dagger_{j,\sigma}$ creates an electron with spin $\sigma$ at site $j$; $\hat{n}_j=\sum_\sigma c^\dagger_{j,\sigma}c_{j,\sigma}$ is the number operator; and $U$ is the on-site Hubbard interaction energy. The bare dispersion is $\epsilon_k=-2t_{\rm el}\cos k$, where the nearest-neighbor hopping $t_{\rm el}\equiv 1$ sets the energy unit.  The operators $b^\dagger_j$ create optical phonons at sites $j$ with dispersion $\Omega(q)=\Omega+2t_{\rm ph}\cos q$ for $\Omega\gg|t_{\rm ph}|$.  The last term is the Holstein electron-phonon coupling~\cite{holstein59}, with effective dimensionless coupling strength $\lambda=g^2/[2t_{\rm el}\sqrt{\Omega^2-4t_{\rm ph}^2}]$~\cite{berciu2013}.

We study this model using Variational Exact Diagonalization (VED) for $N_e=2$ (corresponding to vanishing electron concentration $n=N_e/N\to 0$ for large $N$) and the density matrix renormalization group (DMRG) for finite concentrations $n$.  Technical details for both these well-established  methods are provided in the Supplemental Material.
 
\textit{Results:} We first consider the pure Hubbard model ($\lambda=0$), varying $U$ from attractive values, where the 1D ground state is always paired, to weakly repulsive values, where it is unpaired. The left column of Fig.~\ref{FIG2} shows VED results for $D(\omega_1,\omega_2)$ at $N_e=2$.  Panels (a) and (b), for attractive $U$, show weight confined to the line $\omega_1+\omega_2=2\mu$ [Eq.~(\ref{global})], with a gap along the hypotenuse of the right triangle of leg $\Delta$, consistent with Fig.~\ref{FIG1}(b).  (The finite broadening $\eta$ in Eq.~(\ref{Mkk}) accounts for the small spillover into the forbidden region.)  

This is expected: for $N_e=2$ the ground state contains a single pair, so type-(i) processes are absent.  Furthermore, with a unique final state (the vacuum) after two-electron emission, weight is confined to the diagonal $\omega_1+\omega_2=E_{\bf 0}^{(0,2)}-E_{\bf 0}^{(0,0)}=2\mu$.  In both cases, the maximum intensity occurs at $\omega_1=\mu$ or $\omega_2=\mu$ (solid red lines), allowing $\Delta$ to be read off directly.  Intensity decreases along the diagonal away from these maxima and spreads over a wider energy range for larger $|U|$.

This behavior is determined by the ground-state pair wavefunction $|\psi_{\bf 0}^{(0,2)}\rangle
 = \sum_{\bf k}\phi({\bf k})
   c^\dagger_{{\bf k}\uparrow}c^\dagger_{-{\bf k}\downarrow}|0\rangle$, where $\phi({\bf k})$ is strongly peaked at ${\bf k}=0$ and extends up to $|{\bf k}|\sim 1/\xi$, with $\xi$ the pair size.  The most probable first emission is of an electron with ${\bf k}=0$, leaving the remaining electron in the single-particle ground state (energy cost $\mu$).  Emission of electrons with $|{\bf k}|>0$ is less probable and leaves a higher-energy intermediate state, shifting the associated intensity below $\mu$.  Greater spread of the 2eARPES intensity thus reflects larger contributions from $|{\bf k}|>0$ states, signaling a more compact pair in real space.  Since the quasiparticle dispersion $E_{\bf k}^{(0,1)}$ is accessible via standard ARPES, the energy spread of the 2eARPES signal can be mapped to a momentum spread, providing an estimate of $\xi$. Panel (c) shows VED results for weakly repulsive $U$.  This ground state consists of two unpaired quasiparticles, well approximated by setting $\phi({\bf k})=\delta_{{\bf k},0}$.  Combined with the Pauli exclusion principle, which forbids $\omega_1=\omega_2$, this places the 2eARPES weight near the upper corner of the allowed square [Fig.~\ref{FIG1}(a)].
 
DMRG results at finite $n=0.25$ are shown in panels (d)--(f) of Fig.~\ref{FIG2}.  Panels (d) and (e), for attractive $U$, reproduce weight along the $\omega_1+\omega_2=2\mu$ line (minus the forbidden triangle), further broadened by contributions from pairs with different momenta.  At finite $n$, type-(i) contributions also appear within their allowed region, consistent with Fig.~\ref{FIG1}(b).  As discussed above, this weight concentrates near the corner of the allowed square; however, since the pairs are bosons, Pauli exclusion does not apply and weight is permitted along $\omega_1=\omega_2$.  The values of $\Delta$ extracted from these maps are in excellent agreement with the analytical result $\Delta=\sqrt{U^2+4t_{\rm el}^2}-2t_{\rm el}$~\cite{SawatzkyPRL}. For repulsive $U$, panel (f) shows weight confined to the square of Fig.~\ref{FIG1}(a), consistent with panel (c) but broadened by contributions from the many quasiparticles of the finite-density ground state.
 
We note that Ref.~\cite{Devereaux_2023} presents analogous results for the two-dimensional Hubbard model, confirming that the bounds of Fig.~\ref{FIG1} are independent of dimensionality.
 
We next consider a more realistic model in which pairing is mediated by boson exchange, to identify possible spectroscopic signatures of the pairing glue.  Figure~\ref{FIG3} shows results for the Holstein model $(U=0)$ with Einstein phonons; results for the Hubbard-Holstein model with dispersive phonons are presented in the Supplemental Material.  We focus on the paired ground state with $U=0$, $\Omega=1$, $\lambda=1$, for which the ground state is a liquid of weakly interacting bipolarons~\cite{Klemen2025}.  Panel (a) shows VED results for $N_e=2$, where both electrons are ejected from the single ground-state bipolaron.  Weight appears not only along $\omega_1+\omega_2=2\mu$, as in Figs.~\ref{FIG2}(a),(b), but also along the satellite lines $\omega_1+\omega_2=2\mu-p\Omega$ for $p\geq 1$.  This occurs because the Holstein model admits final states with $p\geq 1$ residual phonons after both electrons are ejected, unlike the Hubbard model where the only final state is the vacuum.  The appearance of these phonon replicas within the type-(ii) wings [Fig.~\ref{FIG1}(b)] directly evidences boson-mediated pairing and allows the boson energy to be read off directly.
 
The intensity is considerably more spread along these diagonals than in Fig.~\ref{FIG2} because bipolaron electrons occupy all momenta (the phonon cloud carrying the remainder), whereas in an attractive Hubbard pair the electrons are restricted to $k_1=-k_2$ near $k=0$. Additional modulations of intensity along both axes, also governed by $\Omega$, reflect the richer intermediate-state spectrum, which includes not only a polaron carrying the transferred momentum but also polaron-plus-phonon continua and higher-energy features.
 
Panels (b)--(d) of Fig.~\ref{FIG3} show DMRG results for increasing concentrations $n$, consistent with expectations.  The most prominent change relative to panel (a) is the growth of type-(i) weight inside the square bounded by the red solid lines.

\textit{Conclusions:} We have demonstrated that the $2\gamma\!\rightarrow\!2e$ 2eARPES configuration provides a direct and unambiguous probe of electronic pairing.  The central result is a universal, model-independent energy separation between type-(i) and type-(ii) contributions to the two-electron coincidence spectrum, producing qualitatively distinct spectral patterns for paired versus unpaired ground states that are immediately recognizable without sophisticated analysis.  Large-scale numerical simulations of the 1D Hubbard and Hubbard-Holstein models confirm these universal bounds and reveal additional information encoded in the spectrum: the distribution of intensity along the type-(ii) diagonal encodes the pair wavefunction and thereby the pair size $\xi$, while in phonon-mediated systems satellite diagonals at $\omega_1+\omega_2=2\mu-p\Omega$ ($p\geq 1$) directly evidence the pairing boson and yield its energy.

These results establish $2\gamma\!\rightarrow\!2e$ 2eARPES as a uniquely powerful spectroscopic probe of paired quantum states, with direct implications for the identification and characterization of pairing in cuprate superconductors and other unconventional superconductors, as well as systems of preformed pairs, where the nature of the paired state and the identity of the pairing glue remain central open questions.

\textit{Acknowledgments:} We thank Steef Smit and Giorgio Levy for useful discussions. J.B. acknowledges the support by the program No. P1-0044 of the Slovenian Research Agency (ARIS) and project KTTK21: {\it Quantum technologies for transportation and communications in the 21st century} supported by  by the University of Ljubljana.  This project was supported in part by the Max Planck--UBC--UTokyo Center for Quantum Materials, the Canada
First Research Excellence Fund (Quantum Materials and Future
Technologies Program), the Natural Sciences and Engineering Research Council of Canada (A.D.\ and M.B.), the Canada Research Chairs Program (A.D.), and the CIFAR Quantum Materials Program (A.D.).
 
\bibliography{manuDomom}

% ---------------------------------------------------------------
% END MATTER
% ---------------------------------------------------------------
%\section*{Universal bounds for $\gamma\!\rightarrow\!2e$ 2e-ARPES}

\section*{End Matter}

\begin{figure}[b]
\includegraphics[width=\columnwidth]{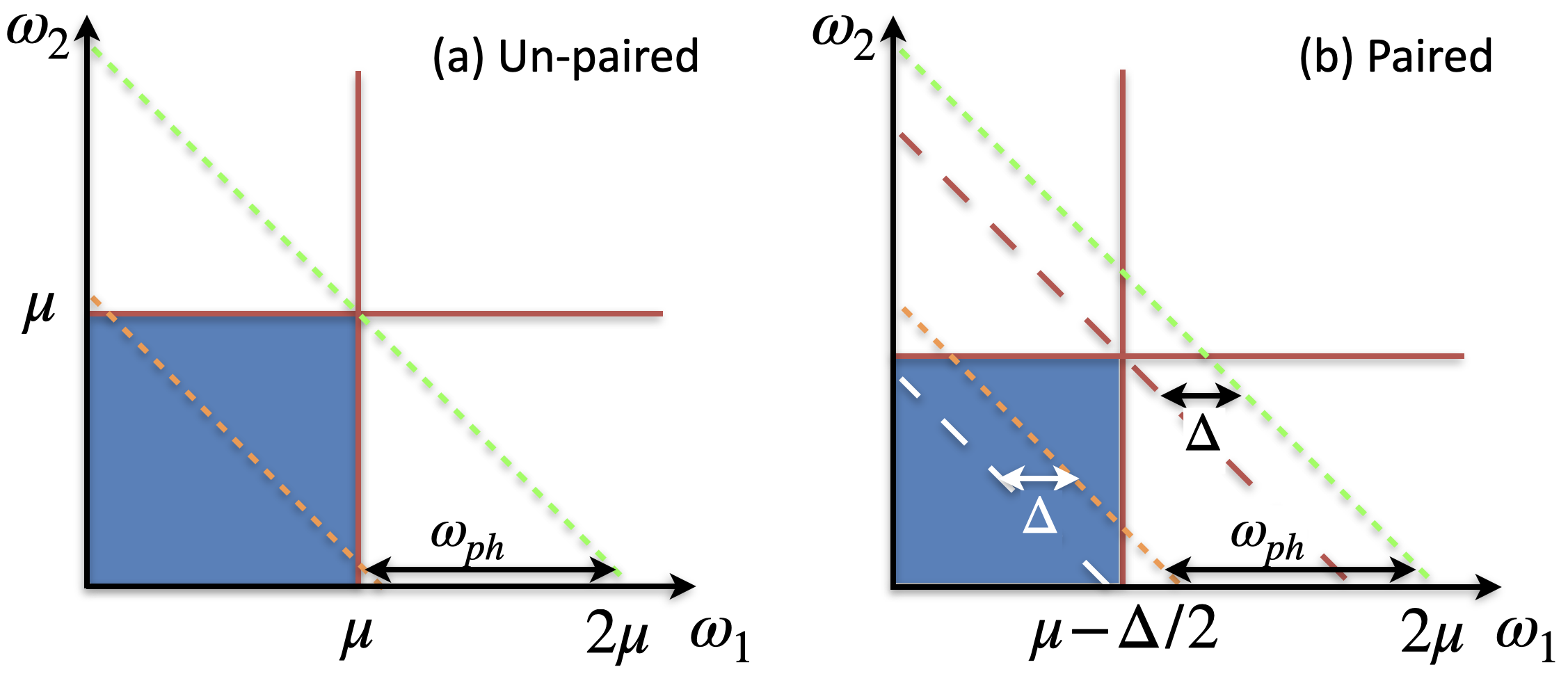}
\caption{\label{S1}
  Counterparts to Fig.~\ref{FIG1} for the $\gamma\!\rightarrow\!2e$  setup.  (a)~Unpaired case: the dotted orange line marks the  global energy maximum for $\gamma\!\rightarrow\!2e$ processes. (b)~Paired case: the dotted orange (dashed white) line marks the  global maximum for type-(ii) [type-(i)] $\gamma\!\rightarrow\!2e$  processes.  Blue shaded regions show where conventional $\gamma\!\rightarrow\!e$ ARPES weight is expected; unlike for the $2\gamma\!\rightarrow\!2e$ setup, here the 2eARPES signal is not  energetically separated from this background.}
\end{figure}

Here we briefly contrast the universal bounds for the $2\gamma\!\rightarrow\!2e$ setup [Fig.~\ref{FIG1}] with their counterparts for the $\gamma\!\rightarrow\!2e$ configuration [Fig.~\ref{S1}].  We retain the notation of the main text: $\omega_i={\bf K}_i^2/2m+W-\omega_{\rm ph}$, $i=1,2$.  The global constraint $\omega_1+\omega_2\leq 2\mu$ for the $2\gamma\!\rightarrow\!2e$ process (dotted green line) becomes $\omega_1+\omega_2\leq 2\mu-\omega_{\rm ph}$ for the $\gamma\!\rightarrow\!2e$ process (dotted orange line), reflecting the reduction by one photon energy in the available initial energy.
 
In the unpaired case [Fig.~\ref{S1}(a)], only type-(i) $\gamma\!\rightarrow\!2e$ processes are possible, but their signal overlaps with conventional $\gamma\!\rightarrow\!e$ ARPES, which is confined to the same blue region $\omega_1\leq\mu$, $\omega_2\leq\mu$ as discussed in Fig.~\ref{FIG1}.  
%Depending on the value of $\omega_{\rm ph}$ relative to $\mu$, some or all of the $\gamma\!\rightarrow\!2e$ weight is obscured by this conventional ARPES background.  
While the two contributions can in principle be separated by their different dependencies on laser intensity (linear versus quadratic), this is experimentally more demanding.
 
In the paired case, the global energy constraints for type-(ii) (orange dotted line) and type-(i) (dashed white line) processes remain separated by $\Delta$, reflecting the different number of broken pairs.  Each diagonal lies an energy $\omega_{\rm ph}$ below its $2\gamma\!\rightarrow\!2e$ counterpart (dotted green and dashed red lines, respectively).  Conventional ARPES intensity (blue shaded region) again obscures all or the most significant part of this region, depending on $\omega_{\rm ph}$.  This spectral overlap with the conventional ARPES signal is the primary reason why we believe that the $\gamma\!\rightarrow\!2e$ configuration is less suited to diagnosing pairing.

\end{document}